\documentclass[12pt]{article}

\usepackage{psfig,amsfonts,apalike}

\newsymbol\lesssim 132E
\newsymbol\gtrsim 1326

\begin{document}

\title{\bf A quantitative approximation scheme for the traveling wave
solutions in the Hodgkin-Huxley model}

\author{C. B. Muratov \\ Department of Mathematical Sciences, \\ New
Jersey Institute of Technology, \\ University Heights, Newark, NJ
07102}

\date\today

\maketitle


\begin{abstract}

We introduce an approximation scheme for the Hodgkin-Huxley model of
nerve conductance which allows to calculate both the speed of the
traveling pulses and their shape in quantitative agreement with the
solutions of the model. We demonstrate that the reduced problem for
the front of the traveling pulse admits a unique solution. We obtain
an explicit analytical expression for the speed of the pulses which is
valid with good accuracy in a wide range of the parameters.

\end{abstract}


\section{Introduction}

Understanding the mechanisms of the propagation of the nerve activity
is one of the fundamental problems in biophysics. The simplest example
of such a propagation is a single solitary traveling pulse of action
potential in an axon \cite{katz}. Today it is well established that
the changes of the membrane potential in nerve tissue are the result
of the complex dynamics of the ionic currents through
voltage-sensitive channels \cite{katz}. The first detailed
quantitative measurements of the ionic currents were performed by
Hodgkin and Huxley in the early 50-s \cite{hh:jp52:500}. Using the
voltage clamp technique, they were able to measure the kinetics of
$\mathrm{Na}^+$ and $\mathrm K^+$ currents in the squid giant
axon. This led them to a set of differential equations which describe
the dynamics of the action potential. Furthermore, by combining these
equations with the cable equation for spreading of current in the axon
they were able to calculate the shape and velocity of the propagating
action potentials \cite{hh:jp52:500,huxley59}. The predictions of
their model turned out to be in a remarkably good agreement with the
experimental observations.

The reason that the model introduced by Hodgkin and Huxley (the HH
model) admits quantitative comparisons with the experiments is that it
contains detailed information about the voltage-dependent kinetics of
the $\mathrm{Na}^+$ and $\mathrm K^+$ channels. Naturally, this makes
the models quite complex and intractable analytically. So far, the
basic tool for studying the HH model were numerical simulations. Note
that because of its complexity, it was not until recently, with the
advent of very fast computers, that the simulations of the HH model
could be done routinely. Even then, one is still required to do
simulations for each set of the parameters of the model. Therefore,
analytical studies which would give functional dependences of the main
parameters of the action potentials on the parameters of the model are
still highly desirable.

The early analytical works on the HH model relied on the strong
separation of the time scales of the (fast) activation and (slow)
inactivation processes. These studies made an assumption that the
$\mathrm{Na}^+$ activation is the fastest process and can be
eliminated adiabatically, what amounts to assuming that the sodium
activation variable $m = m_\infty(V)$, where $m_\infty(V)$ is the
resting value of $m$ at a given membrane voltage $V$
\cite{fitz,casten75,carpenter77,carpenter79}. This leads to a
cubic-like nonlinearity in the equation for the membrane potential. By
further assuming that the $\mathrm{Na}^+$ inactivation and $\mathrm
K^+$ dynamics are much slower than the $\mathrm{Na}^+$ activation, the
problem of the action potential propagation reduces to a single
reaction-diffusion equation for the front of the action potential
\cite{casten75}. A number of simpler models (FitzHugh-Nagumo type)
with similar properties had been introduced to mimic the behavior of
the membrane \cite{fitz,nagumo,rinzel73,casten75,jones91}. The latter
in fact became quite popular for explaining traveling waves phenomena
in a variety of excitable systems in physics, chemistry, and biology
\cite{vasiliev,murray,mikhailov,ko:book}.

Although this kind of analysis leads to a qualitative explanation of
the excitability of the nerve membrane, it fails to give any
quantitative predictions for the speed of the propagating action
potentials. It also predicts that the traveling wave should have the
form of a broad excitation region with the sharp front and back. This
is in contrast to the observations in which the pulse is a narrow
localized region of excitation (a spike). The reason for this is that
in reality it is typically the membrane potential rather than the
$\mathrm{Na}^+$ activation that is the fastest process. For example,
in the squid giant axon the time constant of the membrane potential
change is $\tau_V \sim 0.01 \mathrm{~ms}$, whereas the time constant
of the $\mathrm{Na}^+$ activation is roughly $\tau_{\mathrm{Na}} \sim
0.2 \mathrm{~ms}$. So, the FitzHugh-Nagumo type models are in fact not
adequate for any quantitative predictions of the characteristics of
the action potential. Also, they only qualitatively reveal the
mechanism of the wave propagation.

In the present paper, we introduce an approximation scheme that does
take into account this relationship between the time scales. We will
construct an approximate solution for a single traveling pulse in the
HH model that is in the quantitative agreement with the solutions of
the full HH model. We will investigate the structure of the front of
the traveling pulse and show that it is substantially different from
the conventional case of the FitzHugh-Nagumo type models. We will also
obtain an explicit analytical expression for the speed of the pulses
which agrees with the results of the simulations of the HH model
within 20\% accuracy in a wide parameter range. Using the obtained
solutions, we will construct an approximate solution for the entire
pulse that is also in quantitative agreement with the solutions of the
full HH model.

\section{The Hodgkin-Huxley model}

In the following, we will use the version of the HH model which was
originally introduced by Hodgkin and Huxley to study the behavior of
the squid giant axon \cite{hh:jp52:500} and later adopted by many
researchers as a benchmark of the models of nerve activity. Namely, we
will consider the following equations
\begin{eqnarray} 
C {\partial V \over \partial t} & = & {a \over 2 \rho} {\partial^2 V
\over \partial x^2} + g_{\mathrm{Na}} m^3 h (V_{\mathrm{Na}} - V) +
g_{\mathrm K} n^4 (V_{\mathrm K} - V) + g_l (V_l - V), \label{hh:V} \\
{\partial m \over \partial t} & = & \alpha_m(V) (1 - m) - \beta_m(V)
m, \label{hh:m} \\ {\partial h \over \partial t} & = & \alpha_h(V) (1
- h) - \beta_h(V) h, \label{hh:h} \\ {\partial n \over \partial t} & =
& \alpha_n(V) (1 - n) - \beta_n(V) n.
\label{hh:n}
\end{eqnarray}
Here, as usual, Eq.~(\ref{hh:V}) is the cable equation for the
membrane potential $V$, with $C = 1 ~\mu \mathrm F/\mathrm{cm}^2$ the
membrane capacitance per unit area, $a = 238 ~\mu \mathrm m$ is the
radius of the axon and $\rho = 35.4~\Omega \times \mathrm{cm}$ the
resistivity of the intracellular space; $g_{\mathrm{Na}} = 120
~\mathrm m\Omega^{-1}/\mathrm{cm}^2$, $g_{\mathrm K} = 36 ~\mathrm
m\Omega^{-1}/\mathrm{cm}^2$ are the conductances of the open
$\mathrm{Na}^+$ and $\mathrm K^+$ channels per unit area;
$V_{\mathrm{Na}} = 115 ~\mathrm{mV}$ and $V_{\mathrm K} = -12
~\mathrm{mV}$ are the equilibrium potentials of $\mathrm{Na}^+$ and
$\mathrm K^+$, and $g_l = 0.3 ~\mathrm m\Omega^{-1}/\mathrm{cm}^2$ and
$V_l = 10.5989 ~\mathrm{mV}$ are the leakage conductance per unit area
and the leakage voltage, respectively. With these definitions the
resting potential $V_r = 0$. Similarly, $m$ and $h$ are the activation
and the inactivation variables for the $\mathrm{Na}^+$ channels,
respectively; $n$ is the $\mathrm K^+$ inactivation variable; the
rates $\alpha_{m,h,n}$ and $\beta_{m,h,n}$ as functions of $V$ at
temperature $T = 6.3^\mathrm o~$C can be found in \cite{hh:jp52:500}
(note that \cite{hh:jp52:500} have an opposite sign convention for
$V$). The temperature changes are accounted for by a factor $\phi =
3^{(T - 6.3)/10}$ multiplying all $\alpha$'s and $\beta$'s, $T$ is in
degrees Celsius. The lengths are measured in centimeters and the times
in milliseconds. The voltage $V$ is measured in millivolts.

Equations (\ref{hh:V}) -- (\ref{hh:n}) constitute a closed system of
partial differential equations that {\em quantitatively} describe the
changes in the membrane as functions of time and space. Let us
emphasize that their ingredients are obtained by measurements and
fitting of the parameters to the actual experiments. So, it is
important to understand the relationships between the characteristic
parameters, namely the time scales, in this system. From the
functional form of $\alpha$'s and $\beta$'s \cite{hh:jp52:500} we can
make the following estimates for the time constants $\tau_{m,h,n}$ for
$m, h$, and $n$, respectively, at $T = 6.3^\mathrm o~$C:
\begin{eqnarray}
\tau_m & \sim & 0.2~\mathrm{msec}, \\
\tau_h & \sim & 5 ~\mathrm{msec}, \\
\tau_n & \sim & 3 ~\mathrm{msec}.
\end{eqnarray}
Also, from Eq.~(\ref{hh:V}) one gets the following estimate for the
time scale $\tau_V$ of the variation of the voltage, assuming that all
the $\mathrm{Na}^+$ channels are open:
\begin{eqnarray}
\tau_V \sim C / g_{\mathrm{Na}} \sim 0.01 ~\mathrm{msec}. 
\end{eqnarray}
One can see that the following hierarchy of time scales holds in the
system:
\begin{eqnarray} \label{hier}
\tau_V \ll \tau_m \ll \tau_h, \tau_n.
\end{eqnarray}
The first inequality holds better for sufficiently low temperatures,
and remains qualitatively correct up to the temperatures $T \sim
30^\mathrm o~$C at which the pulses fail to propagate in the HH model
\cite{huxley59}. As we already pointed out in the Introduction, this
is an important property of the system which is not taken into account
in most of the analytical studies of the HH model. In the following,
we will use this hierarchy of time scales to introduce the
approximation scheme for the traveling pulses in this model.

Another important point about the HH model is the fact that the very
nonlinearities in Eq.~(\ref{hh:V}), namely, the powers with which the
variables $m, h$, and $n$ enter the equation are determined
experimentally \cite{hh:jp52:500}. Furthermore, these powers
correspond to the number of particles involved in the operation of the
respective channels and therefore represent significant physical
quantities. As will be seen below, these powers in fact play crucial
role in our studies.

Before going to the analysis of the traveling pulses, let us discuss
the basic physics of the excitability in the HH model. In the rest
state, the $\mathrm{Na}^+$ channels are basically closed, at $V = 0$
the equilibrium values for $m$ and $h$ are $m_0 \simeq 0.05$ and $h_0
\simeq 0.60$, respectively, while the $\mathrm K^+$ channels are
partially open, with $n_0 \simeq 0.32$. If, by applying an external
stimulus, the membrane voltage $V$ is increased to $\sim 10
~\mathrm{mV}$, the $\mathrm{Na}^+$ channels will start opening on the
time scale of order $\tau_m$. The influx of the $\mathrm{Na}^+$ ions
will in turn lead to the increase of the membrane potential $V$ on the
time scale intermediate between $\tau_V$ and $\tau_m$ (see below),
resulting in the positive feedback. The membrane potential $V$ will
come close to the resting potential $V_\mathrm{Na}$, while the
$\mathrm{Na}^+$ channels will become mostly open with $m \simeq
1$. During this time, the slow inactivation variables $h$ and $n$ will
remain almost unchanged. After that, the slow inactivation variables
$h$ and $n$ will start to react, closing the $\mathrm{Na}^+$ and
opening the $\mathrm K^+$ channels, which will drive the potential $V$
back to equilibrium. In the spatially extended system the diffusive
spreading of the current in front of the excitation region in the axon
will provide the sustaining force for the propagation of the pulse
along the axon. In that sense, from the physical point of view the
traveling pulse in the nerve axon is a classical example of an {\em
autosoliton} --- self-sustained solitary inhomogeneous state in an
active dissipative system whose existence is determined only by the
nonlinearities of the system and not the initial conditions
\cite{ko:book}.

\section{Solitary pulse}

We are now going to construct an approximate traveling wave solution
in the form of a solitary pulse, using the ideas introduced in the
preceding section. Let us introduce a self-similar variable $z = x - c
t$, where $c$ is the propagation speed of the pulse. Then,
Eqs.~(\ref{hh:V}) -- (\ref{hh:n}) for a traveling wave with speed $c$
will become
\begin{eqnarray}
&& {a \over 2 \rho} {d^2 V \over dz^2} + c C {d V \over dz} +
g_{\mathrm{Na}} m^3 h (V_{\mathrm{Na}} - V) + g_{\mathrm K} n^4
(V_{\mathrm K} - V) + g_l (V_l - V) = 0, \nonumber \\ \label{hz:V} \\
&& c {d m \over d z} + \alpha_m(V) (1 - m) - \beta_m(V) m = 0,
\label{hz:m} \\ && c {d h \over d z} + \alpha_h(V) (1 - h) -
\beta_h(V) h = 0, \label{hz:h} \\ && c {d n \over d z} + \alpha_n(V)
(1 - n) - \beta_n(V) n = 0. \label{hz:n}
\end{eqnarray}
The boundary conditions for these equations are
\begin{eqnarray}
V(\pm\infty) = 0, ~~~m(\pm\infty) = m_0, ~~~ h(\pm\infty) = h_0, ~~~
n(\pm\infty) = n_0,
\end{eqnarray}
where $m_0, h_0$, and $n_0$ are the values of $m$, $n$, and $h$ in the
rest state, respectively. For the chosen functions $\alpha$ and
$\beta$ the rest state $V = 0$ is unique and stable. 

\begin{figure}
\centerline{\psfig{figure=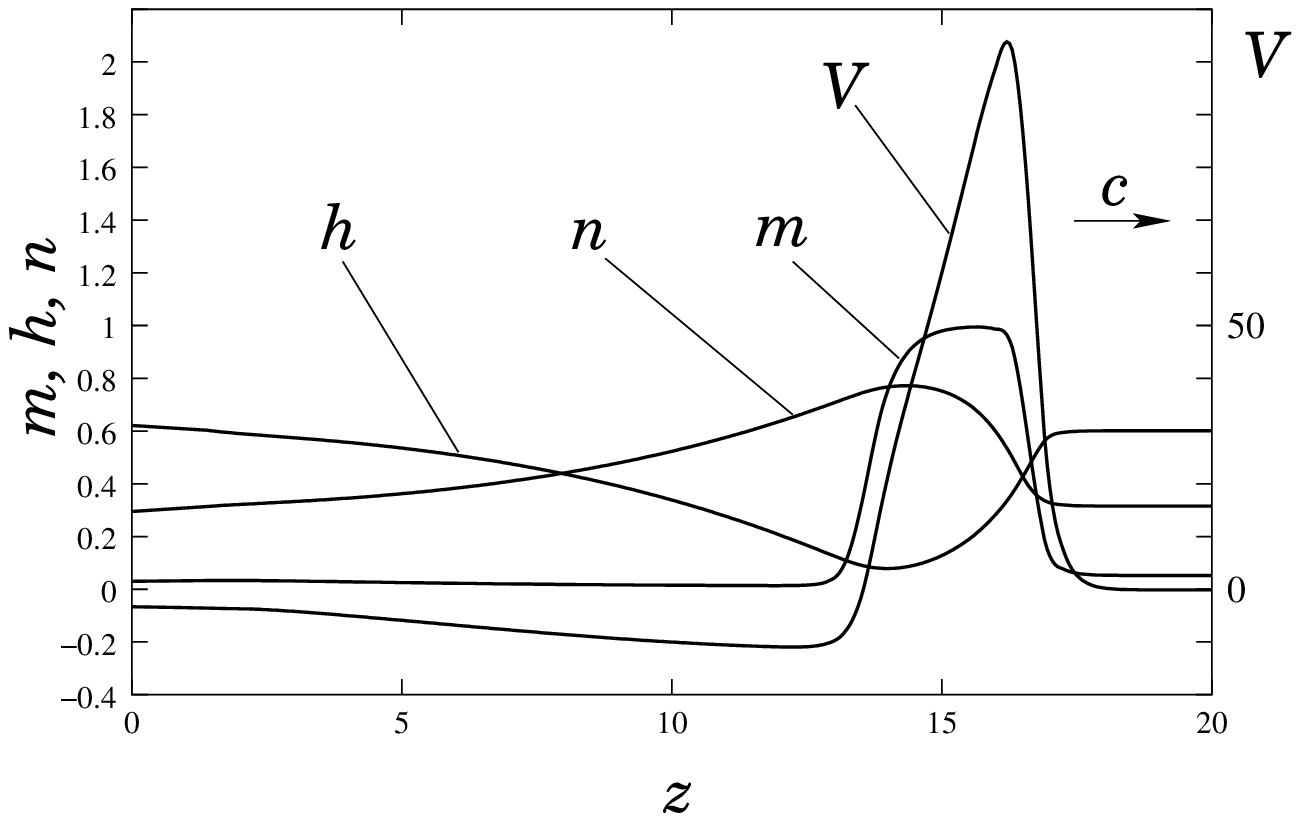,width=3.25in}}
\caption{The numerical solution of Eqs.~(\ref{hz:V}) -- (\ref{hz:n})
at $T = 6.3^\mathrm o$ C.} \label{f:hh}
\end{figure}

The solution of Eqs.~(\ref{hz:V}) -- (\ref{hz:n}) in the form of a
traveling solitary pulse obtained numerically at the ``standard''
temperature $T = 6.3^\mathrm o$ C is shown in Fig. \ref{f:hh}.
From this figure one can see several features of the solution which we
will use in the approximation scheme that we are going to
construct. First, observe that the length scale of the rise of the
potential is substantially smaller than that of the fall of the
potential. Second, during the rise of the potential the variables $h$
and $n$ remain almost unchanged at their resting values $h_0$ and
$n_0$. Third, in front of the spike the value of $m$ (that is, $m_0$)
is practically zero.

Let us use the above facts to simplify Eqs.~(\ref{hz:V}) --
(\ref{hz:n}). Since the values of $h$ and $n$ change little in the
front of the spike, we may replace them by their values $h_0$ and
$n_0$ at rest and disregard Eqs~(\ref{hz:h}) and
(\ref{hz:n}). Furthermore, since the value of $m_0$ is very small,
with very good accuracy we may assume it to be zero. Therefore, in the
rest state we will have $g_\mathrm{K} n_0^4 V_\mathrm{K} + g_l V_l =
0$ with very good accuracy, so these terms drop out of
Eq.~(\ref{hz:V}). Also, the coefficient multiplying $-V$ in the last
two terms of Eq.~(\ref{hz:V}) is of order $0.7$ and is much smaller
than the contribution from the term $g_{\mathrm{Na}} m^3 h$ during the
rise of the potential when $m$ is not close to zero, so these terms
can be dropped as well. What we are then left with is a an equation
for $V$ coupled only to the equation for $m$ with a number of terms
dropped. Observe that since $V$ is much faster than $m$ when $m \sim
1$, the value of $m$ has to be sufficiently small in order for the
nontrivial collective dynamics of $V$ and $m$ to be possible. This
allows to further simplify Eq.~(\ref{hz:m}) by neglecting the terms
proportional to $m$. After making all these approximations, we are
left with the following set of equations
\begin{eqnarray} 
&& {a \over 2 \rho} {d^2 V \over dz^2} + c C {dV \over dz} +
g_\mathrm{Na} m^3 h_0 (V_\mathrm{Na} - V) = 0, \label{h:V} \\ && c {d
m \over dz} + \alpha_m(V) - \alpha_m(0) = 0, \label{h:m}
\end{eqnarray}
instead of Eqs.~(\ref{hz:V}) and (\ref{hz:m}). Note that we added a
term $- \alpha_m(0)$ to Eq.~(\ref{h:m}) in order for this equation to
be consistent with the approximate boundary conditions ahead of the
spike front
\begin{eqnarray} \label{appbc}
m(+\infty) = 0, ~~~V(+\infty) = 0, ~~~V_z(+\infty) = 0,
\end{eqnarray}
where $V_z = d V / dz$. We can do this in our approximation scheme
since the value of $\alpha_m(0)$ is in practice very small compared to
$\alpha_m(V_\mathrm{Na})$.

Let us assume that the characteristic value of $m$ in the front is
$\bar m \ll 1$ and the characteristic width of the front is $l$. Then,
since all the terms in Eqs.~(\ref{h:V}) and (\ref{h:m}) should be of
the same order of magnitude, we obtain the following estimates
\begin{eqnarray} \label{scale}
{a \over \rho l^2} \sim {c C \over l} \sim g_\mathrm{Na} h_0 \bar m^3,
~~~ {c \bar m \over l} \sim \alpha_m(V_\mathrm{Na}).
\end{eqnarray}
From these one can also estimate the characteristic time scale for the
rise of the potential in the pulse as $\tau \sim l / c \sim (C /
\bar\alpha_m^3 g_\mathrm{Na} h_0)^{1/4}$, where
\begin{eqnarray}
\bar\alpha_m = \alpha_m(V_\mathrm{Na}) - \alpha_m(0),
\end{eqnarray}
so
\begin{eqnarray} \label{tau}
\tau \sim \left( \tau_V \tau_m^3 \right)^{1/4}.
\end{eqnarray}
One can see from this equation that the dynamics in the front of the
traveling pulse will indeed occur on the time scale intermediate
between $\tau_V$ and $\tau_m$.

From the estimates above we immediately conclude that in the traveling
spike
\begin{eqnarray} \label{bar}
c = \bar c \left( {\bar\alpha_m^3 a^4 g_\mathrm{Na} h_0 \over 16
\rho^4 C^5} \right)^{1/8}, ~~\bar m = \left( {\bar\alpha_m C \over
g_\mathrm{Na} h_0} \right)^{1/4}, ~~ l = \left( { a^4 \over 16 \rho^4
C^3 \bar\alpha_m^3 g_\mathrm{Na} h_0} \right)^{1/8},
\end{eqnarray}
where $\bar c$ is a constant of order 1. Substituting the parameters
of the HH model at $T = 6.3^\mathrm o$ C, we see that $\bar m \simeq
0.6$, what corresponds to the relevant quantity $\bar m^3 \simeq 0.2$,
which is indeed rather small.

\begin{figure}
\centerline{\psfig{figure=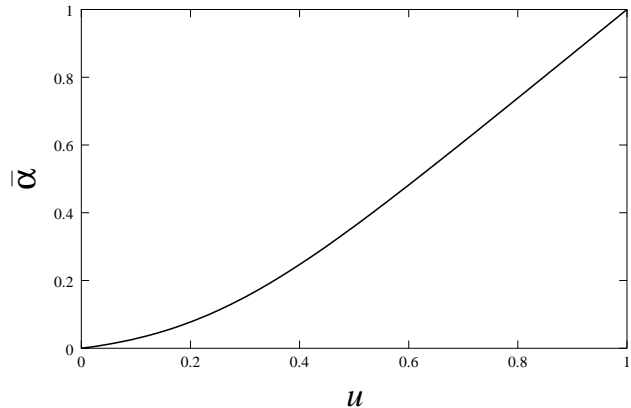,width=3.25in}}
\caption{The dependence $\bar\alpha(u)$.}
\label{f:ab}
\end{figure}

Indeed, let us introduce the following new variables:
\begin{eqnarray}
\xi = {z \over l}, ~~~s = {\bar c^2 m \over \bar m}, ~~~u = {V \over
V_\mathrm{Na}}, ~~~v = \bar\alpha(u) {du \over ds}, \label{new}
\end{eqnarray}
where
\begin{eqnarray}
\bar\alpha(u) = {\alpha_m(V_\mathrm{Na} u) - \alpha_m(0) \over
\alpha_m(V_\mathrm{Na}) - \alpha_m(0)}.
\end{eqnarray}
The latter is plotted in Fig. \ref{f:ab}.
Note that this way the time scale in Eq.~(\ref{h:m}) has been absorbed
into the constant $\bar\alpha_m$.

In terms of the variables introduced in Eq.~(\ref{new}) and after a
few manipulations one can rewrite Eqs.~(\ref{h:V}) and (\ref{h:m}) in
the following form 
\begin{eqnarray}
\bar\alpha(u) {du \over ds} & = & v, \label{u} \\ \bar\alpha(u) {dv
\over ds} & = & v - {s^3 (1 - u) \over \bar c^8}, \label{v} \\
\bar\alpha(u) {d\xi \over ds} & = & - {1 \over \bar c}, \label{xi}
\end{eqnarray}
where now $s$ is an independent variable. These transformations can be
done for $0 < u < 1$ since $\bar\alpha(u)$ is always positive for $u
\not = 0$ (see Fig. \ref{f:ab}). Note that these equations do not have
any $\xi$ dependence in their right-hand side, so it suffices to solve
Eqs.~(\ref{u}) and (\ref{v}) only. The solution for $\xi(s)$ can then
be obtained by a simple integration.

The problem now became substantially simpler because instead of
solving the nonlinear boundary value problem for Eqs.~(\ref{hz:V}) --
(\ref{hz:n}), one now needs to solve the initial value problem for
Eqs.~(\ref{u}) and (\ref{v}). Indeed, according to Eq. (\ref{appbc}),
when $z \rightarrow +\infty$ we have $m \rightarrow 0$, so $s
\rightarrow 0$ as $\xi \rightarrow +\infty$. This means that $u = 0$
and $v = 0$ at $s = 0$, since also $d u /d \xi = - \bar c v
\rightarrow 0$ as $\xi \rightarrow +\infty$ [see Eqs.~(\ref{appbc}),
(\ref{u}), and (\ref{xi})]. One should though be careful to specify
what exactly happens near $s = 0$ since $\bar\alpha(0) = 0$. To do
this, let us divide Eq.~(\ref{v}) by Eq.~(\ref{u}). We get
\begin{eqnarray} \label{phase}
{dv \over du} = 1 - {s^3 (1-u) \over \bar c^8 v}.
\end{eqnarray}
When $s \rightarrow 0$, we have $dv/du \rightarrow 1$, provided that
$v(s)$ has a non-zero derivative at $s = 0$ (the latter follows from
the physical considerations). Therefore, according to Eqs.~(\ref{u})
and (\ref{v}), as $s \rightarrow 0$, the solution will behave like
\begin{eqnarray} \label{uv:asym}
u(s) = {s \over \bar\alpha'(0)} + o(s), ~~~~~v(s) = {s \over
\bar\alpha'(0)} + o(s),
\end{eqnarray}
where the prime means differentiation. Here we expanded
$\bar\alpha(u)$ in the neighborhood of zero and took into account that
$\bar\alpha' (0) \not = 0$.

It is not difficult to see from Eq.~(\ref{phase}) that for $0 < u < 1$
and $v > 0$ the projection of the phase trajectory on the $u-v$ plane
will lie below the line $u = v$. Since $du/ds > 0$ and there are no
fixed points in this region of $u$ and $v$, the solution $u(s), v(s)$
will cross either the line $u = 1$ or $v = 0$. By direct inspection of
Eqs.~(\ref{u}) and (\ref{v}) one can see that this intersection is
transversal. Observe that the intersection of the lines $u = 1$ and $v
= 0$ is a fixed point in this plane.

According to Eqs.~(\ref{u}) and (\ref{v}), once the solution leaves
the region bounded by the lines $u = v$, $u = 1$ and $v = 0$, it can
never come back to this region. Indeed, if the solution crosses the
line $u = 1$ at some $v > 0$ in the $u-v$ plane, it will then have
$dv/ds > 0$ for all $s$, so $v$ will stay positive. If on the other
hand, the solution crosses the line $v = 0$ at some $u < 1$, we will
have both $dv/ds < 0$ and $du/ds < 0$ afterwards. In fact, it is easy
to see that the only trajectory on which $u$ and $v$ will remain
bounded for all $s$ is the one which connects the point $u = 0, v = 0$
with $u = 1, v = 0$. Therefore, this trajectory is precisely the one
that corresponds to the front of the traveling pulse.

\begin{figure}
\centerline{\psfig{figure=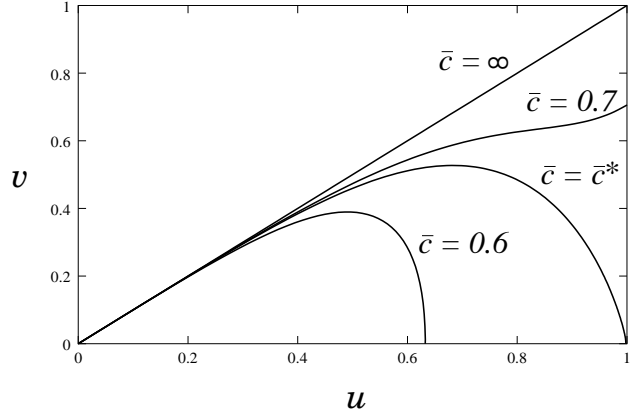,width=3.25in}}
\caption{The numerical solution of Eqs.~(\ref{u}) and (\ref{v}) in the
$u-v$ plane $v(u)$ at different $\bar c$.}
\label{f:c}
\end{figure}

Of course, not all the speeds $\bar c$ will produce this kind of
trajectory. It is clear that if $\bar c$ is very large, the trajectory
will cross the line $u = 1$ in the $u-v$ plane at $v$ close to 1. On
the other hand, if $\bar c$ is very small, the trajectory will cross
the line $v = 0$ at very small $u$. Figure \ref{f:c} shows the results
of the numerical solution of Eqs.~(\ref{u}) and (\ref{v}) at different
values of $\bar c$.
From this numerical solution we found that the trajectory that
connects $u = 0, v = 0$ and $u = 1, v = 1$ exists only for a unique
value of $\bar c = \bar c^*$.

In fact, it is possible to prove that such a trajectory indeed exists
and is unique at a unieque value of $c$. To do this, let us see what
happens with the trajectory as the value of $\bar c$ is changed. For
convenience, we will use $u$ instead of $s$ as an independent
variable. Let $v_0(u)$ and $s_0(u)$ be a trajectory in the region
bounded by $u = v$, $u = 1$ and $v = 0$ with the initial conditions
$v_0(0) = 0, s_0(0) = 0$ for some $\bar c = \bar c_0$. To calculate
the changes in the trajectory $\delta v(u), \delta s(u)$ as $\bar c$
is changed by $\delta \bar c$, we write the equations in variations
for $\delta v$ and $\delta s$ obtained from Eqs.~(\ref{u}) and
(\ref{v})
\begin{eqnarray}
{d \over du} \delta s & = & - {\bar\alpha(u) \over v_0^2} \delta v,
\label{ds} \\
{d \over du} \delta v & = & {s_0^3 (1 - u) \over \bar c_0^8 v_0^2}
\delta v - {3 s_0^2 (1 - u) \over \bar c_0^8 v_0} \delta s + {8 s_0^3
(1 - u) \over \bar c_0^9 v_0} \delta\bar c. \label{dv}
\end{eqnarray}
Since the change in $\bar c$ does not affect the initial conditions,
we should have
\begin{eqnarray}
\delta v(0) = 0, ~~~~~\delta s(0) = 0.
\end{eqnarray}
Note that according to Eqs.~(\ref{uv:asym}) we have $v_0 \sim u$ and
$s_0 \sim u$, so $\delta s \sim u^3$ and $\delta v \sim u^3$ in the
neighborhood of $u = 0$.

According to Eq.~(\ref{dv}), when $u \rightarrow 0$ we have ${d \over
du} \delta v > 0$ for $\delta\bar c > 0$, so $\delta v > 0$. In turn,
according to Eq.~(\ref{ds}), ${d \over du} \delta s < 0$ and therefore
$\delta s < 0$. This means that the derivatives ${d \over du} \delta
v$ and ${d \over du} \delta s$ do not change signs, so $\delta v > 0$
everywhere for $\delta \bar c > 0$ and vice versa. Therefore, when the
value of $\bar c$ changes from 0 to $\infty$, the point at which the
trajectory crosses either the line $u = 1$ or the line $v = 0$ in the
$u-v$ plane will go monotonically from $u = 0, v = 0$ to $u = 1, v =
1$. Since it depends continuously on $\bar c$, there is a unique value
of $\bar c = \bar c^*$ at which this point coincides with $u = 1, v =
0$. Numerically, the value of $\bar c^* = \frac{2}{3}$ up to the
fourth digit. Thus, the dynamics in the pulse front uniquely
determines its propagation speed.

\begin{figure}
\centerline{\psfig{figure=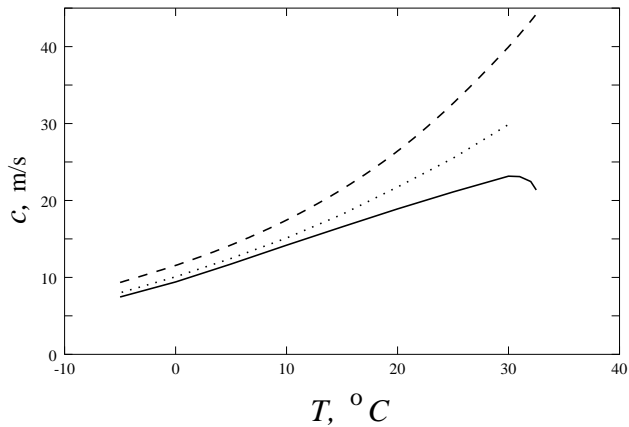,width=3.25in}}
\caption{The speed $c$ of the traveling pulse as a function of
$T$. The solid line shows the results of the numerical solution of the
full HH model. The dashed line is the prediction of Eq.~(\ref{c}). The
dotted line is the result of the solution of the HH model without the
$h$ and $n$ dynamics.}
\label{f:trav}
\end{figure}

\begin{figure}
\centerline{\psfig{figure=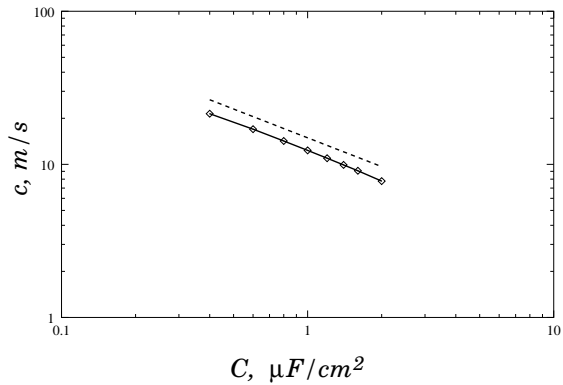,width=3.25in}}
\caption{The speed $c$ of the traveling pulse as a function of the
membrane capacitance $C$. The solid line is the result of the
numerical solution of the HH model, the dashed line is the prediction
of Eq.~(\ref{c}). }
\label{hhtravc}
\end{figure}

Thus, we have obtained an approximate {\em analytical} expression for
the speed of the traveling pulses in the HH model:
\begin{eqnarray} \label{c}
c = \frac{2}{3} \left( {a^4 \bar\alpha_m^3 g_\mathrm{Na} h_0 \over 16
\rho^4 C^5} \right)^{1/8}.
\end{eqnarray}
Equation (\ref{c}) predicts the speed of the traveling pulse as a
function of the parameters. To compare this predicted speed with the
results obtained from the numerical solution of the HH model, we plot
the speed $c$ as a function of temperature obtained from Eq.~(\ref{c})
and from the numerical simulations of the HH model (see also
\cite{huxley59}) in Fig. \ref{f:trav} (recall that the temperature
dependence is contained in the value of $\bar\alpha_m$).
As can be seen from this figure, the approximate expression for the
speed of the pulse given by Eq.~(\ref{c}) agrees with the results for
the HH model within $20 \%$ at temperatures below $15^\mathrm o$ C. We
would like to emphasize that this is the kind of accuracy with which
the HH model {\em itself} predicts the speeds of the traveling pulses
as compared to the experiments. At higher temperatures the agreement
between Eq.~(\ref{c}) and the results of the simulations of the HH
model becomes worse, and at the temperatures of the propagation
threshold Eq.~(\ref{c}) fails completely. We have also checked that
Eq.~(\ref{c}) predicts the correct dependences on the other parameters
with similar accuracy at low enough temperatures. For example,
Fig. \ref{hhtravc} shows a comparison of the prediction of
Eq.~(\ref{c}) with the results of the numerical simulations of the HH
model at $T = 6.3^\mathrm o$ C as the value of the membrane
capacitance $C$ per unit area is varied, on the log-log plot.
Note that the slopes of the two graphs in Fig. \ref{hhtravc} agree
very well with each other. The agreement of the slopes is not as good
for the log-log plot of the dependence of $c$ on $g_\mathrm{Na}$ with
other parameters fixed. This is probably the consequence of the fact
that the errors introdiced by our approximation depend on
$g_\mathrm{Na}$ stronger than the prediction of the approximation $c
\sim g_{Na}^{1/8}$.

Incidentally, if the factor of $2/3$ in Eq.~(\ref{c}) is replaced by
$5/9$, it will give the the speed of the pulse within a few per cent
of that found in the full HH model for $T < 15^\mathrm o$ C. Note that
if one assumes that $m$ is the fastest variable
\cite{fitz,casten75,carpenter77,carpenter79} and calculates the speed
of the traveling wave, one will get the value which is an order of
magnitude greater than the actual value.

Observe that Fig. \ref{f:trav} also shows the dependence of the speed
of the pulse on temperature obtained from the simulations of the HH
model without the $h$ and $n$ dynamics (the dotted line). Note that
the insignificance of these variables is one of the key assumptions in
deriving Eq.~(\ref{c}). One can see that this solution gives an even
better approximation to the speed of the pulse. This problem, however,
cannot be treated analytically in the same manner as that for
Eqs.~(\ref{h:V}) and (\ref{h:m}).

Let us emphasize that the existence of the front solution is
essentially determined by the complicated interplay of the $V$ and $m$
kinetics, so the problem does not reduce to simple phase plane
analysis, in contrast to most studies of the traveling waves in
excitable systems
\cite{fitz,rinzel73,casten75,carpenter77,carpenter79,vasiliev,mikhailov}.
Note that similar situation takes place in a class of excitable
systems in which the so-called spike autosolitons are realized
\cite{om:prl95,mo1:98}. These models also give rise to the complicated
front structures that are similar to the one realized in the HH model.

The validity of the approximations made by us is violated in two
cases. First, when the temperature becomes sufficiently high, the
dynamics of the $m$ variable becomes faster, so the separation of the
time scales $\tau_m$ and $\tau_V$ used in our arguments will no longer
be justified. One of the implications of the absence of this scale
separation is the fact that the characteristic value of $m = \bar m$
in the front can no longer be treated as small. This allows us to
derive a criterion for the validity of our approximations
\begin{eqnarray} \label{c1}
{\bar\alpha_m C \over g_\mathrm{Na} h_0} \lesssim 1,
\end{eqnarray}
which is equivalent to $\bar m \lesssim 1$ [see Eq.~(\ref{bar})]. In
the case of the squid giant axon this criterion shows the
applicability of our results up to $T \simeq 25^\mathrm o$ C, in good
agreement with Fig. \ref{f:trav}.

Another problem may occur when the temperature becomes too low and the
variable $m$ too slow. In this case the effective time scale $\tau$ of
the dynamics in the front of the pulse slows down [see
Eq.~(\ref{tau})], so at some point it may become comparable to the
leakage time scale $\tau_l \sim C/g_l \sim 3$ msec. In this case one
can no longer discard the leakage and the $\mathrm K^+$ contributions
to the membrane current in Eq.~(\ref{hz:V}). Thus, the second validity
criterion becomes [see Eq.~(\ref{scale})]
\begin{eqnarray} \label{c2}
{(g_l + g_\mathrm{K} n_0^4)^4 \over \bar\alpha_m^3 C^3 g_\mathrm{Na}
h_0} \lesssim 1.
\end{eqnarray}
For the squid giant axon, this quantity becomes comparable to 1 only
for the unrealistically low temperatures $T \lesssim -30^\mathrm o$ C.

As can be seen from Eqs.~(\ref{c1}) and (\ref{c2}), the quality of the
approximations used by us should increase with the increase of
$g_\mathrm{Na}$. In fact, our procedure for finding the spike's speed
and the front profile is the leading order of the asymptotic limit
$m_0 \rightarrow 0$ and $g_\mathrm{Na} \rightarrow \infty$.

\begin{figure}
\centerline{\psfig{figure=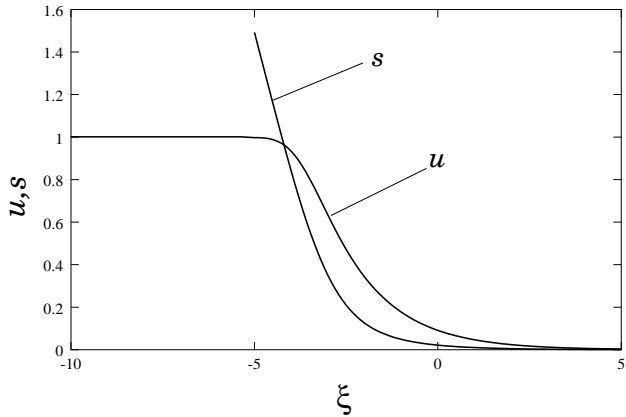,width=3.25in}}
\caption{The functions $u(\xi)$ and $s(\xi)$ obtained numerically from
Eqs.~(\ref{u}) -- (\ref{xi}) for $\bar c = \bar c^*$.}
\label{f:hh2}
\end{figure}

Figure \ref{f:hh2} shows the functions $u(\xi)$ and $s(\xi)$ for $\bar
c = \bar c^*$ obtained numerically.
One can see that $u(\xi)$ has the form of a front connecting the rest
state $u = 0$ with the excited state $u = 1$ which in the original
variables corresponds to the saturation value $V = V_\mathrm{Na}$. 

As can be seen from Fig. \ref{f:hh2}, the distribution $s(\xi)$ behind
the front approaches a straight line with the slope $-\bar c^*$. In
terms of the original variables, this should correspond to the
unlimited growth of $m$ behind the front. This however, should not be
a problem since this happens only when $m \sim \bar m \ll 1$. When $m$
becomes of order 1, the approximations used to derive Eq.~(\ref{xi})
ceases to be valid. On the other hand, when this happens, $V$ should
already be very close to $V_\mathrm{Na}$, so on the time scale $\tau_m
\ll \tau_{h,n}$ the variable $m$ will simply exponentially relax to $m
= m_\infty(V_\mathrm{Na})$ behind the front, where, as usual
\begin{eqnarray} \label{mi}
m_\infty(V) = {\alpha_m(V) \over \alpha_m(V) + \beta_m(V)},
~~~\tau_m(V) = {1 \over \alpha_m(V) + \beta_m(V)}.
\end{eqnarray}
This will happen at distances of order $c \tau_m \gg l$, since $\tau_m
\gg \tau$ [see Eq.~(\ref{tau})]. If we assume that on the time scale
$\tau_m$ the front was located at $z = 0$, the distribution of $m$
that properly matches with that in Fig. \ref{f:hh2} will be
\begin{eqnarray} \label{sh}
m(z) = m_\infty(V_\mathrm{Na}) \{1 - \exp[z/c \tau_m(V_\mathrm{Na}) ]
\}, 
\end{eqnarray}

As was already noted, in the back of the spike and in the refractory
tail the voltage $V$ changes substantially slower than in the
front. Since this happens when $m \sim 1$, the voltage is indeed the
fastest variable, so we can eliminate it adiabatically from the
equations. If we put all the derivatives in Eq.~(\ref{hz:V}) to zero,
we find
\begin{equation} \label{vv}
V = {g_\mathrm{Na} m^3 h V_\mathrm{Na} + g_\mathrm{K} n^4 V_\mathrm{K}
+ g_l V_l \over g_\mathrm{Na} m^3 h + g_\mathrm{K} n^4 + g_l}.
\end{equation}
This expression uniquely determines the value of $V$ as a function of
$m, h$, and $n$.

\begin{figure}
\centerline{\psfig{figure=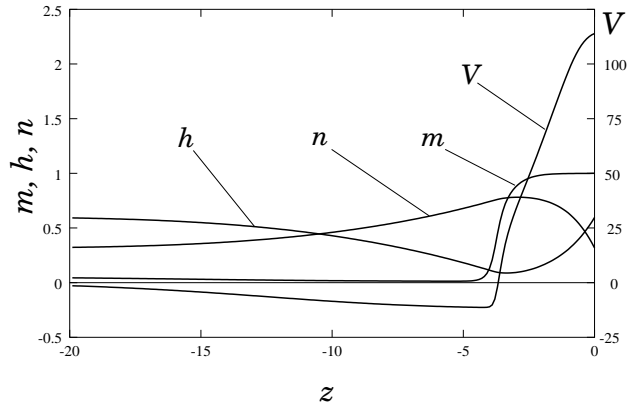,width=3.25in}}
\caption{The numerical solution of Eqs.~(\ref{hz:m}) -- (\ref{hz:n}),
(\ref{vv}), and (\ref{bcslow}).}
\label{f:hh4}
\end{figure}

To find the approximate distributions of all the variables in the back
and behind the spike we simply need to solve the initial value problem
given by Eqs.~(\ref{hz:m}) -- (\ref{hz:n}) with $c$ given by
Eq.~(\ref{c}) and the following initial conditions
\begin{eqnarray} \label{bcslow}
m(0) = 1, ~~~ h(0) = h_0, ~~~n(0) = n_0,
\end{eqnarray}
where we assumed that on the even larger length scale $c \tau_{h,n}$
the front is located at $z = 0$. This initial value problem can be
straightforwardly solved numerically. The result of this solution is
shown in Fig. \ref{f:hh4}
Note that the changes in temperature will only change the
characteristic length of this solution and not its shape.

\begin{figure}
\centerline{\psfig{figure=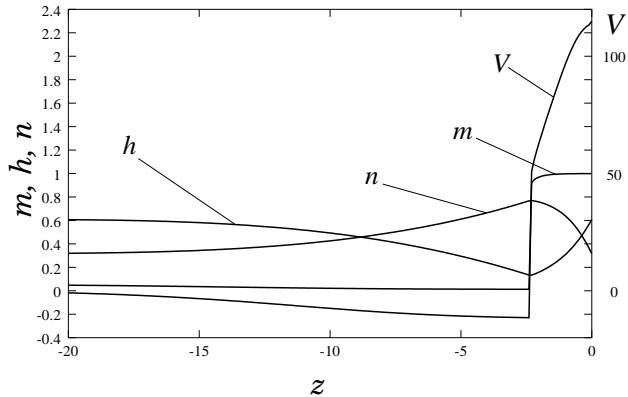,width=3.25in}}
\caption{The numerical solution of Eqs.~(\ref{hz:h}), (\ref{hz:n}),
(\ref{vv}), and (\ref{bcslow}) with $m = m_\infty(V)$ given by
Eq.~(\ref{mi}).}
\label{f:hh7}
\end{figure}

One can simplify the procedure of finding the distributions of $m$,
$h$, and $n$ by using the fact that $\tau_m \ll \tau_{h,n}$ by
adiabatically eliminating $m$. This will amount to replacing $m$ by
$m_\infty$ from Eq.~(\ref{mi}) in Eq.~(\ref{vv}) and then solving for
$V$ as a function of $h$ and $n$. The result of the numerical solution
of Eqs.~(\ref{hz:h}) and (\ref{hz:n}) with these approximations is
shown in Fig. \ref{f:hh7}.
This figure shows a good agreement of the slow variables $h$ and $n$
in the refractory tail. Also, observe an abrupt jump in the back of the
spike. This is due to the fact that now the value of $V$ is not
uniquely determined by $h$ and $n$, so at some point in the solution a
jump occurs from one branch of the dependence $V(h, n)$ to the other
(see also \cite{carpenter77,carpenter79}). Note that while the
adiabatic elimination of $m$ works well for the refractory tail, it
fails in the back of the pulse (compare Figs. \ref{f:hh4} to
\ref{f:hh7}).

\begin{figure}
\centerline{\psfig{figure=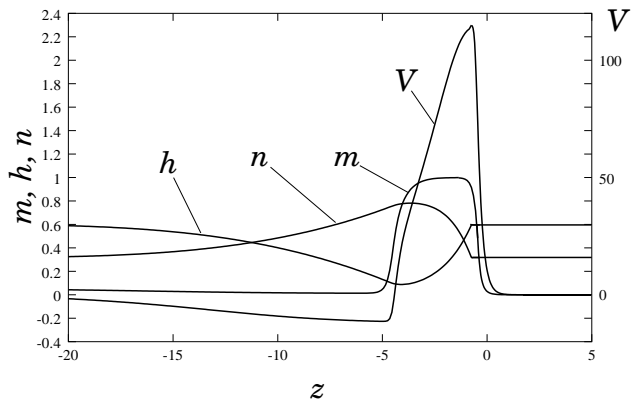,width=3.25in}}
\caption{An approximate solution for the entire pulse.}
\label{f:hh1}
\end{figure}

The results in Figs. \ref{f:hh2}, \ref{f:hh4}, and Eq.~(\ref{sh}) can
be combined to give a quantitative approximation for the whole
pulse. This is done in Fig. \ref{f:hh1} for $T = 6.3^\mathrm o$ C.
One can see remarkable similarity between the solution of the full HH
model shown in Fig. \ref{f:hh} and the one shown in
Fig. \ref{f:hh1}. Thus, our approximation scheme has been able to
capture {\em quantitatively} the essential features of the traveling
pulses in the HH model.

\section{Conclusions}

We have developed a method that allows to approximately compute the
shape and the parameters of the traveling spikes in the HH model of
conductance along an axon. Our method is different from the
conventional approach
\cite{fitz,rinzel73,casten75,carpenter77,carpenter79} in the fact that
it treats the {\em membrane potential} rather than the sodium
activation variable as the fast variable. We show that this is in fact
the case for the typical set of the parameters of the Hodgkin-Huxley
model. This leads to a good {\em quantitative} agreement between the
predictions of the theory and the results of the numerical simulations
of the HH model.

Let us emphasize that the HH model itself gives only an approximate,
although quite accurate description of the dynamics of the action
potential in an actual axon. What we find is that in a broad range of
the parameters the approximation introduced by us gives an error which
is in fact comparable to the error produced by the HH model itself as
opposed to the experiments. For example, at $T = 18.5 ^\mathrm o $C
the speed of the pulse in the squid giant axon was found to be 21.2
m/s \cite{hh:jp52:500}. The direct numerical simulation of the HH
model produces the speed of 18.8 m/s, while our procedure which for
this temperature is already near the limit of its applicability gives
24.7 m/s. Therefore, we suggest that the ideas of our analysis can in
fact be built into simpler and more tractable models of nerve
conductance which will yet be able to give quantitative agreement with
the experimental observations.

One of the observations from the analysis made by us is the fact that
the speed of the traveling spikes in the HH model depends very weakly
on the slow state variables of the membrane. Indeed, according to
Eq.~(\ref{c}), the speed of the spike is independent of the value of
$n$ in front of the spike and is proportional to $h_0^{1/8}$, so a
change by a factor of 2 in $h_0$ will result in only a 10\% change in
the speed. This makes a perfect biological sense. Thus, propagation of
the nerve pulses is indeed a very robust phenomenon.

Another observation one can make from Eq.~(\ref{c}) is that if one
assumes that in addition to the membrane conductance $C$ there is an
extra capacitance associated with each sodium channel, there exists a
density of the channels at which the speed is maximal
\cite{hodgkin75}. Indeed, let us assume that $C = C_0 + N
C_\mathrm{Na}^*$ and $g_\mathrm{Na} = N g_\mathrm{Na}^*$, where $C_0$
is the capacitance of the membrane without the channels, $N$ is the
channel density, $C_\mathrm{Na}^*$ is the capacitance associated with
a single channel, and $g_\mathrm{Na}^*$ is the maximum conductance of
a single channel. For the squid giant axon these parameters are
estimated to be $C_0 = 0.8 ~\mu \mathrm {F/cm}^2$, $C_\mathrm{Na}^* =
4 \times 10^{-18} ~\mathrm{F}$, $g_\mathrm{Na}^* = 2.4 \times 10^{-12}
~\Omega^{-1}$, and $N = 500 ~\mu\mathrm{m}^{-2}$
\cite{hodgkin75}. Substituting these expressions into Eq.~(\ref{c}),
one gets the speed of the pulse as a function of $N$. It is not
difficult to see that this function has a maximum at $N = N_{max} =
C_0 /( 4 C_\mathrm{Na}^*)$. For the numerical values above we find
$N_{max} = 500~\mu\mathrm{m}^{-2}$, what suggests that the channel
density in the axon is indeed at the optimum level. The fact that we
get exactly the same value of $N$ as the one observed may be a bit
fortuitous because of the approximate nature of Eq.~(\ref{c}). Note
that because of the very slow dependence of the speed on
$g_\mathrm{Na}$, the maximum of the dependence $c(N)$ is in fact very
flat, so a change of $N$ by a factor of 2 from $N_{max}$ results only
in a 7\% difference in $c$.

So far, we were talking only about the traveling wave solutions in the
form of the solitary spikes. It is not difficult to see that our
method can be extended to spike trains as well. Indeed, the speed of a
spike in a spike train is determined by the value of the slow variable
$h$ in front of the spike [see Eq.~(\ref{c})], which, however, is now
different from the equilibrium value $h_0$ and must be
determined. Outside of the spike fronts one has to solve the equations
of the slow dynamics given by Eqs.~(\ref{hz:m}) -- (\ref{hz:n}) in
which the fast variable $V$ has been eliminated adiabatically via
Eq.~(\ref{vv}). These equations have to be solved as an initial value
problem for $-L \leq z \leq 0$ with $m(0) = m_\infty(V_\mathrm{Na})$,
$h(0) = h_s$, and $n(0) = n_s$. Here $h_s$ and $n_s$ are the values of
$h$ and $n$ in the spike, respectively, $L$ is the spatial period of
the spike train, and we assumed that the front of one of the spikes in
the spike train is located at $z = 0$. The spikes are also assumed to
travel to the right with the speed given by Eq.~(\ref{c}) in which
$h_0$ is replaced by $h_s$. Then, the values of $h_s$ and $n_s$ have
to be found self-consistently from the condition that $h(-L) = h_s$
and $n(-L) = n_s$.

We implemented this procedure numerically. To find the values of $h_s$
and $n_s$ as functions of $L$, we started with a reasonable initial
guess for $h_s$ and $n_s$ and then solved the initial value problem up
to $z = -L$. The values of $h$ and $n$ at $z = -L$ were then taken as
the new values for $h_s$ and $n_s$, respectively, and the procedure
was iterated. We found a fast convergence to the periodic
solution. Then, from the value of $h_s$ we obtained the speed of the
spike train as a function of $L$ and therefore the dependence of the
spike frequency on the period. Our main finding was that in the region
of the applicability of our approximation [Eq.~(\ref{c1})] the speed
of the spike trains is practically {\em independent} of the period, so
they have almost no dispersion. This also makes good biological
sense. In fact, the amount of the dispersion we found is comparable to
the error introduced by our approximation scheme. Since we are only
interested in quantitative predictions, we do not present these
results in detail here. Also, our method fails for periods $L \lesssim
10$ cm, for which a substantial amount of dispersion was found in the
simulations of the full HH model \cite{miller81}. Nevertheless, let us
point out that the results obtained with our method are in a good
qualitative agreement with those obtained for the full HH model
\cite{miller81}. In particular, according to our numerical solution
outlined above there exists a period of the spike train for which the
speed of the spikes reaches maximum, greater than that of the solitary
spike due to a slight overshoot of the $h$ variable behind the
spike. However, the magnitude of this overshoot is so small that it
only changes the speed of the spike by a fraction of a per cent. So,
for practical purposes the spike trains with period $L \gtrsim 10$ cm
may be considered dispersionless.

In short, we have introduced an approximation scheme which allows to
make quantitative predictions of the shape and the parameters of the
traveling pulses in the HH model of nerve conductance. We hope that
our results will provide an easy and convenient tool for analyzing the
fascinating complexity of the neural activity.

\section{Acknowledgments}
The author gratefully acknowledges many valuable discussions with
C. Peskin and J. Rinzel.

\bibliographystyle{apalike}

\bibliography{../main}

\end{document}